\documentclass[letterpaper,12pt]{JHEP3}
\usepackage{amsfonts, amsmath, amssymb, amsthm, cite}
\usepackage[english]{babel}

\newcommand{\eq}{\begin{equation}}
\newcommand{\feq}{\end{equation}}
\newcommand{\eqn}{\begin{eqnarray}}
\newcommand{\feqn}{\end{eqnarray}}

\title{Nonextremal black holes in gauged supergravity and the real formulation of
special geometry}

\author{Dietmar Klemm$^{ab}$ and Owen Vaughan$^{cd}$ \\
$^a$ Dipartimento di Fisica, Universit\`a di Milano, \\
\hspace*{0.15cm} Via Celoria 16, 20133 Milano, Italy. \\
$^b$ INFN, Sezione di Milano, Via Celoria 16, 20133 Milano, Italy. \\
$^c$ Department of Mathematical Sciences, \\
\hspace*{0.15cm} University of Liverpool, Peach Street, Liverpool L69 7ZL, UK. \\
$^d$ Department of Mathematics and Center for Mathematical Physics, \\
\hspace*{0.15cm} University of Hamburg, Bundesstrasse 55, 20146 Hamburg, Germany.
}
\preprint{IFUM-999-FT\\ LTH 951}

\abstract{We give a rather general recipe for constructing nonextremal black hole solutions to
${\cal N}=2$, $D=4$ gauged supergravity coupled to abelian vector multiplets. This problem
simplifies considerably if one uses the formalism developed in \cite{Mohaupt:2011aa}, based on dimensional
reduction and the real formulation of special geometry. We use this to find new nonextremal black
holes for several choices of the prepotential, that generalize the BPS solutions found in
\cite{Cacciatori:2009iz}. Some physical properties of these black holes are also discussed.
}


\begin{document}

\section{Introduction}

Black holes in anti-de~Sitter (AdS) spaces provide an important testground to address fundamental
questions of quantum gravity like holography. These ideas originally emerged from string theory,
but became then interesting in their own right, for instance in recent applications to condensed matter
physics (cf.~\cite{Hartnoll:2009sz} for a review), where black holes play again an essential role,
since they provide the dual description of certain condensed matter systems at nonzero temperature.
A basic ingredient of realistic condensed matter systems is the presence of a finite density of charge
carriers, which implies the necessity of a bulk $\text{U}(1)$ gauge field. A further step in modeling
strongly coupled holographic systems is to include the leading relevant (scalar) operator in the
dynamics. This is generically uncharged, and is dual to a neutral scalar field in the bulk.
We are thus naturally led to consider nonextremal charged black holes in gauged supergravity
with scalar fields turned on. Unfortunately, there are not many known solutions of this type\footnote{The
most notable exceptions are perhaps the four-charge black holes in the stu model \cite{Duff:1999gh}
and the rotating solutions of \cite{Chong:2004na,Chow:2010sf,Chow:2010fw}.},
and to remedy this is one of the scopes of our paper.

BPS black holes by definition admit Killing spinors, and thus satisfy first-order equations,
which facilitates enormously their construction. In particular, the classification of all supersymmetric
backgrounds of ${\cal N}=2$, $D=4$ matter-coupled gauged supergravity \cite{Cacciatori:2008ek,Klemm:2009uw,Klemm:2010mc,Meessen:2012sr} provides a systematic method to obtain BPS solutions, without
the need to guess some suitable ans\"atze. This has led to some surprising results, for instance the
construction of genuine supersymmetric black holes with spherical symmetry in the stu
model \cite{Cacciatori:2009iz}. A crucial ingredient for the existence of these solutions is the presence of
nonconstant scalar fields. These black holes were then further studied and generalized
in \cite{Dall'Agata:2010gj,Hristov:2010ri,Klemm:2011xw,Colleoni:2012jq}\footnote{For related work on
this subject see also \cite{Sabra:1999ux,Hristov:2010eu}.}.

On the other hand, once we go away from the BPS case to nonextremality, we have to solve the full
second-order field equations, which is a formidable task. Nevertheless, we will see below that,
guided by the supersymmetric case and by the consideration of known nonextremal solutions
in minimal gauged supergravity, it is possible to give a rather general recipe for constructing
nonextremal black hole solutions to matter-coupled gauged supergravity as well. A crucial ingredient is a formalism in which the second-order equations of motion take a particularly simple form.

In \cite{Mohaupt:2011aa} a new formulation of dimensional reduction of ${\cal N}=2$, $D=4$ ungauged supergravity (the c-map) was presented, which made extensive use of a real, symplectically covariant, formulation of special geometry. A crucial step in the procedure was to absorb a metric degree of freedom into the moduli fields in order to lift a hypersurface constraint, an idea that was first developed in \cite{Mohaupt:2009iq} for $D=5$. The three-dimensional Lagrangian then takes a remarkably simple form, and solutions to the full second-order equations of motion can be found easily and naturally. Since these solutions lift to stationary solutions of the $D = 4$ theory, the procedure provides a powerful solution generating technique for the original theory. One requires neither spherical symmetry nor staticity, and it is applicable to completely generic target manifolds (i.e. not restricted to symmetric spaces). For the $D = 5$ case this approach has been used to find non-extremal black hole solutions \cite{Mohaupt:2010fk}, and for a review of both the $D = 4$ and $D = 5$ cases see \cite{Mohaupt:2011ab}. A similar technique was also considered in \cite{Meessen:2011aa} for static, spherically symmetric backgrounds, and has been used to find $D = 5$ black string solutions \cite{Meessen:2012su}.

It is clear that the formalism developed in \cite{Mohaupt:2011aa} can be immediately adapted to the case of Fayet-Iliopoulos gauged supergravity, as the bosonic part of the Lagrangian only gets modified by a potential term (which is unaffected by dimensional reduction). Since the three-dimensional metric can be completely general it naturally allows for asymptotically AdS solutions. An added bonus is that the Fayet-Iliopoulos potential can be expressed in a particularly simple way in terms of the Hesse potential\footnote{One must take care to distinguish between the Fayet-Iliopoulos potential $V$ and the Hesse potential $H$.}, which is a function that appears in the real formulation of special geometry and replaces the role of the holomorphic prepotential.

The remainder of this paper is organized as follows: In section \ref{N=2sugra} we briefly review
${\cal N}=2$, $D=4$ Fayet-Iliopoulos gauged supergravity and the formalism developed in
\cite{Mohaupt:2011aa}.
Section \ref{nonextr} contains the construction of new nonextremal black holes for various prepotentials
together with a discussion of some of their physical properties.
We conclude in \ref{conclusions} with some final remarks.
The appendices contain a derivation of the Hesse potentials for the models under consideration as
well as a formula for the scalar potential in terms of the Hesse potential.

\section{${\cal N} = 2$ gauged supergravity}
\label{N=2sugra}

Our starting point is the Lagrangian of ${\cal N} = 2$ Fayet-Iliopoulos gauged supergravity coupled to $n$ Abelian vector multiplets, where the gauging is with respect to a $\text{U}(1)$ subgroup of the
$\text{SU}(2)$ R-symmetry group. The bosonic part of the Lagrangian is given by
\begin{eqnarray}
		{\cal L}_4 &\sim& -\tfrac{1}{2} R_4 
-  g_{IJ} \partial_{\hat{\mu}} X^I \partial^{\hat{\mu}} \bar{X}{}^J  \nonumber \\
&&\qquad + \tfrac{1}{4} {\cal I}_{IJ} F^I_{\hat{\mu} \hat{\nu}} F^{J \hat{\mu} \hat{\nu}} + \tfrac{1}{4} {\cal R}_{IJ} F^I_{\hat{\mu} \hat{\nu}} \tilde{F}^{J \hat{\mu} \hat{\nu}}  - V\left(X,\bar{X}\right) \;. \nonumber
\end{eqnarray}
Here, and throughout this paper, we use the conventions of \cite{Mohaupt:2011aa}. The equations of motion without the potential term are covariant with respect to $\text{Sp}(2n + 2, \mathbb{R})$ duality transformations.
In order to obtain physical fields, one must impose that the complex moduli fields $X^I$ are subject to two real constraints that fix dilatations and $\text{U}(1)$ phase transformations. The dilatations are fixed by imposing
the D-gauge 
\[
	\bar{X}NX = -1 \;.
\]
The $\text{U}(1)$ transformations, corresponding to the overall phase of the $X^I$, can be fixed by imposing any appropriate condition, such as $\text{Im}( X^0) = 0$. However, following \cite{Mohaupt:2011aa} we will not impose any $\text{U}(1)$ gauge fixing condition until after dimensional reduction. 

The potential term in the Lagrangian is given by
\[
	V\left(X,\bar{X}\right) = -2 g_I g_J \left({\cal I}^{IJ} + 8 \bar{X}^I X^J\right) \;,
\]
where the $g_I$ denote the gauge coupling constants. Using the D-gauge, we can write this as
\[
	V\left(X,\bar{X}\right) = -2 g_I g_J \left({\cal I}^{IJ} - \frac{8}{\bar{X}NX} \bar{X}^I X^J\right) \;.
\]
It is more convenient to use this form of the potential as it is now a homogeneous function of degree zero. The potential is not covariant with respect to $\text{Sp}(2n + 2, \mathbb{R})$ duality transformations, and
therefore breaks the covariance of the equations of motion.


\subsection{Dimensional reduction and the real formulation of special geometry}

	If we impose that backgrounds are stationary then we can integrate out the redundant timelike direction and obtain a three-dimensional effective Lagrangian. This is what is meant by dimensional reduction over time. The four-dimensional metric is decomposed as
	\[
		ds^2_4 = - e^{\phi} \left( dt + V_\mu dx^\mu \right)^2 + e^{-\phi} g_{3\mu\nu} dx^\mu dx^\nu\,.
	\]
	We now follow the procedure outlined in \cite{Mohaupt:2011aa}. After dimensional reduction, and certain field redefinitions, we obtain the following three-dimensional effective Lagrangian:
	\begin{align}
			{\cal L}_3 \;\sim\;&  - \tfrac{1}{2} R_3 - \tilde{H}_{ab} \left(\partial_\mu q^a \partial^\mu q^b -  \partial_\mu \hat{q}^a \partial^\mu \hat{q}^b \right)   + \frac{1}{2H} V(q)\nonumber \\ 
			&- \frac{1}{H^2} \left( q^a \Omega_{ab} \partial_\mu q^b \right)^2 + \frac{2}{H^2} \left( q^a \Omega_{ab} \partial_\mu \hat{q}^b \right)^2  \nonumber \\
			&- \frac{1}{4 H^2} \left( \partial_\mu \tilde{\phi} + 2\hat{q}^a \Omega_{ab} \partial_\mu \hat{q}^b  \right)^2 \;, \label{eq:3dLag} 
 	\end{align}
	where $\mu, \nu = 1,2,3$ and $a,b,c = 0, \ldots, 2n + 1$, and the matrix $\tilde{H}_{ab}$ depends only on the non-hatted fields, $\tilde{H}_{ab} = \tilde{H}_{ab}(q)$. Without the potential term, the equations of motion and the Lagrangian itself are covariant with respect to the full $\text{Sp}(2n + 2,\mathbb{R})$ duality transformation group. 
	
	Let us take a moment to explain the origin of the fields appearing in the three-dimensional Lagrangian, whereby providing a dictionary to the standard fields of ${\cal N} = 2$ supergravity. The fields $q^a = (x^I,y_I)^T$, where $I = 0, \ldots, n$, represent the degrees of freedom descending from both the complex scalar fields $X^I,\bar{X}^I$ and the KK-scalar $e^{\phi}$, and are given explicitly by
	\begin{equation} \label{xy}
	\begin{aligned}
		x^I &:= \text{Re}\, Y^I = e^{\phi/2} \text{Re}\, X^I  \;,  \\
		y_I &:= \text{Re}\, F_I(Y) = e^{\phi/2} \text{Re}\, F_I(X)  \;.
	\end{aligned}
	\end{equation}
	The intermediate complex coordinates $Y^I$ are simply a rescaling of the original coordinates through the expression $Y^I = e^{\phi/2} X^I$. One can understand this as allowing the original scalars $X^I$, which are constrained by the D-gauge, to absorb the KK-scalar and become unconstrained fields, which we denote by $Y^I$. The KK-scalar can then be interpreted as a dependent field,
	\[
		e^{\phi(Y,\bar{Y})} = -\bar{Y}NY \;.
	\]
	The Lagrangian is still invariant under the $\text{U}(1)$ symmetry, which acts as a local phase transformation of the rescaled coordinates $Y^I$. 
	
	The scalar fields $\hat{q}^a = (\frac{1}{2} \zeta^I, \frac{1}{2} \tilde{\zeta}_I)^T$ descend from the degrees of freedom of the gauge fields, as is most easily seen via their derivatives
	\begin{equation} \label{eq:field_strengths}
	\begin{aligned}
		\partial_\mu \zeta^I &:= F_{\mu 0}^{I}  \;, \\
		\partial_\mu \tilde{\zeta}_I &:= G_{I|\mu 0} \;.
	\end{aligned}
	\end{equation}
	The scalar field $\tilde{\phi}$ represents the degree of freedom descending from the KK-vector $V_\mu$.
	The constant matrix $\Omega$ is given by 
	\[
		\Omega_{ab} = \left( \begin{array}{cc} 0 &	\mathbb{I}	\\ -\mathbb{I} & 0 \end{array} \right) \;,
	\]
	and represents the canonical symplectic form associated with the $q^a$ coordinates, which are, in particular, Darboux coordinates. 
	
	The function $H$ is the Hesse potential of the conical affine special K\"ahler manifold associated with the scalar moduli space \cite{Freed:1997dp}. It is proportional to the Legendre transformation of the imaginary part of the holomorphic prepotential \cite{Cortes:2001}, and also happens to be proportional to the K\"ahler potential,
 	\begin{equation}
 		H(x,y) = 2\text{Im} F(Y(x,y)) - 2y_I u^I(x,y) = \tfrac{1}{2} \bar{Y} N Y  = -\tfrac{1}{2}e^\phi \;.
 		\label{eq:Hesse_KK}
 	\end{equation}
 	In this expression one observes the identification of the Hesse potential with the KK-scalar. This is because the field redefinition $Y^I = e^{\phi/2} X^I$ causes the KK-scalar to be absorbed by the complex scalar fields.
 	The matrix $\tilde{H}_{ab}$ is defined in terms of the Hesse potential by
 	\[
 		\tilde{H}_{ab} := \frac{\partial^2}{\partial q^a \partial q^b} \tilde{H} \;, \qquad \text{where} \qquad
 		\tilde{H} := -\tfrac{1}{2} \log \left( -2H \right) \;.
 	\] 
 	The Hesse potential plays a distinguished role when formulating special geometry in terms of special real coordinates. It replaces the role of the holomorphic prepotential in the sense that it completely determines the dynamics of the Lagrangian.

\subsection{Static backgrounds}

	Let us now specialise further to static backgrounds. In this case the KK-vector vanishes, which corresponds in these coordinates to 
	\begin{equation}
		\partial_\mu \tilde{\phi} + 2\hat{q}^a \Omega_{ab} \partial_\mu \hat{q}^b = 0 \;. \label{static}
	\end{equation}
	We will also make the ansatz
	\begin{equation}
		q^a \Omega_{ab} \partial_\mu q^b = q^a \Omega_{ab} \partial_\mu \hat{q}^b = 0 \;,
		\label{additional-ansatz}
	\end{equation}
	which greatly simplifies the equations of motion, and is a natural ansatz in static backgrounds. This ansatz is automatically satisfied for the solutions found in \cite{Cacciatori:2009iz}. The effective Lagrangian in this case is given by the first line of (\ref{eq:3dLag}),
	\begin{align}
			{\cal L}_3 \;\sim\;&  - \tfrac{1}{2} R_3 - \tilde{H}_{ab} \left(\partial_\mu q^a \partial^\mu q^b - \partial_\mu \hat{q}^a \partial^\mu \hat{q}^b \right) + \frac{1}{2H} V(q) \;. \label{eq:3dLag_static} 
 	\end{align}

	The condition (\ref{additional-ansatz}) explicitly breaks the local $\text{U}(1)$ covariance of the Lagrangian and equations of motion, as it relates the $q^a$ coordinates, which transform under local $\text{U}(1)$ transformations, with the $\hat{q}^a$ coordinates, which do not. It can therefore be seen as a gauge fixing condition for the $\text{U}(1)$ isometry. 

\subsection{Equations of motion}

	It is straight-forward to work out the equations of motion from the three-dimensional static Lagrangian (\ref{eq:3dLag_static}). By varying $q^a,\hat{q}^a$ and $g_{3\mu\nu}$ respectively we find the equations
	\begin{align}
			& \nabla^\mu \left[\tilde{H}_{ab}\partial_\mu q^b \right] -  \tfrac{1}{2}\partial_a \tilde{H}_{bc} \left(\partial_\mu q^b \partial^\mu q^c - \partial_\mu \hat{q}^b \partial^\mu \hat{q}^c \right)  + \partial_a \left( \frac{1}{4H} V(q) \right) = 0 \;, \label{eq:eom1}\\
			&\nabla^\mu \left[\tilde{H}_{ab}\partial_\mu \hat{q}^b \right]  = 0 \;, \label{eq:eom2}\\
			&\tilde{H}_{ab} \left(\partial_\mu q^a \partial_\nu q^b - \partial_\mu \hat{q}^a \partial_\nu \hat{q}^b \right) - \frac{1}{2H} g_{\mu \nu} V(q) = -\tfrac{1}{2}R_{3 \mu \nu}  \;.\label{eq:eom3}
	\end{align}
	The  equation (\ref{eq:eom2}) can be solved immediately to give
	\[
		\tilde{H}_{ab}\partial_\mu \hat{q}^b = \partial_\mu {\cal H}_a \;,
	\]
	where ${\cal H}_a$ are harmonic functions. It is convenient to write the other equations of motion in terms of a natural set of dual coordinates $q_a := -\tilde{H}_{ab} q^b$, which satisfy
	\[
		q_a = \frac{1}{H} \left( \begin{array}{c} -v_I \\ u^I \end{array} \right) = \frac{1}{H} \left( \begin{array}{c} -\text{Im}(F_I) \\ \text{Im}(Y^I) \end{array} \right) \;.
	\]
	In this case the remaining equations of motion can be written simply as
	\begin{align}
			\Delta q_a  +  \tfrac{1}{2}\partial_a \tilde{H}^{bc} \Big(\partial_\mu q_b \partial^\mu q_c - \partial_\mu {\cal H}_b \partial^\mu {\cal H}_c \Big)  + \partial_a \left( \frac{1}{4H} V(q) \right) &= 0 \;, \label{eq:eom_dual1}\\
			\tilde{H}^{ab} \Big(\partial_\mu q_a \partial_\nu q_b - \partial_\mu {\cal H}_a \partial_\nu {\cal H}_b \Big) - \frac{1}{2H} g_{\mu \nu} V(q) &= -\tfrac{1}{2}R_{3 \mu \nu}  \;.\label{eq:eom_dual2}
	\end{align}
	The potential $V(q)$ can be expressed in terms of the Hesse potential through  (\ref{eq:potential_H}).

\subsection{Metric ansatz}

Following \cite{Cacciatori:2009iz} we now make the following ansatz for the 3d metric:
\[
ds_3^2 = dz^2 + e^{2\Phi(z,w,\bar w)} dw d\bar w\,,
\]
where $\Phi$ is separable,
\eq
\Phi(z,w,\bar w) = \psi(z) + \gamma(w,\bar w)\,, \label{Phi-separable}
\feq
and $\gamma$ satisfies the Liouville equation
\eq
\Delta_{(2)}\gamma \equiv 4\partial_w\partial_{\bar w}\gamma = -\kappa e^{2\gamma}\,, \label{liouville}
\feq
with $\kappa$ a constant. \eqref{liouville} means that the two-metric $e^{2\gamma}dwd\bar w$ has
constant curvature. As a solution of \eqref{liouville} we shall take
\eq
e^{2\gamma} = \left[1 + \frac{\kappa}4w\bar w\right]^{-2}\,.
\feq
We also assume that the fields $q^a$ and $\hat{q}^a$ only depend on $z$,
\[
q^a = q^a(z) \;, \hspace{3em} \hat{q}^a = \hat{q}^a(z)\,.
\]


\section{Nonextremal solutions} \label{nonextr}

We shall now construct nonextremal black holes for various choices of the prepotential $F$.
In order to illustrate the general idea, let us first see what happens in minimal gauged
supergravity, where nonextremal solutions are known.
One possible way to obtain this is to consider the stu model with
\begin{equation}
F = -2i(X^0X^1X^2X^3)^{1/2}\,, \label{alternative-stu}
\end{equation}
and set $X^0=X^1=X^2=X^3$, such that $F=-2i(X^0)^2$. This has zero vector multiplets (just the
graviphoton). If we set all $g_I$ equal ($g_I\equiv g/(2\sqrt2)$) as well as $p^I=p^0\,\forall I$,
the black hole solution (3.33) of \cite{Cacciatori:2009iz} reduces to
 \begin{equation}
 ds^2 = -4g^2\left(z-\frac1{2z}\right)^2dt^2 + \frac{dz^2}{g^2\left(z-\frac1{2z}\right)^2} + \frac{z^2}{g^2}
 e^{2\gamma}dwd\bar w\,. \label{metr-minimal}
 \end{equation}
 Note that $\kappa=-1$ in this case in order to have a genuine black hole rather than a naked
 singularity. Introducing new coordinates according to
 \[
 t = \frac{\tau}{2g}\,, \qquad z = gr\,, \qquad w = 2\tanh\frac{\vartheta}2 e^{i\varphi}\,,
 \]
\eqref{metr-minimal} becomes
\begin{equation}
ds^2 = -\left(gr-\frac1{2gr}\right)^2d\tau^2 + \frac{dr^2}{\left(gr-\frac1{2gr}\right)^2} + r^2(d\vartheta^2+
\sinh^2\vartheta d\varphi^2)\,, \label{metr-minimal2}
\end{equation}
and the fluxes (equ.~(3.34) of \cite{Cacciatori:2009iz}) read
\begin{equation}
 F^0=F^1=F^2=F^3=\frac1{2\sqrt2 g}\sinh\vartheta d\vartheta\wedge d\varphi\,. \label{fluxes-minimal}
 \end{equation}
\eqref{metr-minimal2} and \eqref{fluxes-minimal} admit the nonextremal generalization
\begin{eqnarray}
  ds^2 &=& -V(r)d\tau^2 + \frac{dr^2}{V(r)} + r^2(d\vartheta^2+\sinh^2\vartheta d\varphi^2)\,, \label{nonextr-min} \\
  F^I &=& \frac p{\sqrt2}\sinh\vartheta d\vartheta\wedge d\varphi\,, \nonumber
\end{eqnarray}
where 
 \[
 V(r) = -1-\frac mr + g^2r^2+\frac{p^2}{r^2}\,.
 \]
 The BPS solution \eqref{metr-minimal2}, \eqref{fluxes-minimal} corresponds to the special case $m=0$
 ($m$ is the mass parameter), $p=1/(2g)$. Note that \eqref{nonextr-min} can be written in the form
 \begin{equation}
 ds^2 = -Vd\tau^2 + \frac1{Vg^2}(dz^2 + e^{2(\psi+\gamma)}dwd\bar w)\,,
 \end{equation}
 where, in terms of the coordinate $z$,
 \[
 V = -1-\frac{mg}z+z^2+\frac{p^2g^2}{z^2}\,,
 \]
 and
 \begin{equation}
 e^{2\psi} = Vz^2 = -z^2-mgz+z^4+p^2g^2\,.
 \end{equation}
 The key observation is that $e^{2\psi}$ is (like in the BPS case $m=0$, $p=1/(2g)$) still a quartic
 polynomial in $z$, but now there is also a linear term, and the magnetic charge $p$ is no more
 fixed. Note that we are free to include also a cubic term. This corresponds to adding nut charge, but then
 the solution will not be static anymore.
 
As a further example, consider still the prepotential \eqref{alternative-stu}, and set now $X^1=X^2=X^3$,
such that $F=-2i(X^0)^{1/2}(X^1)^{3/2}$. This is the $t^3$ model, which has one vector multiplet. A BPS
black hole solution to this model can be obtained from (3.33), (3.34) of \cite{Cacciatori:2009iz} by taking
$g_1=g_2=g_3$, $p^1=p^2=p^3$, $\alpha^1=\alpha^2=\alpha^3$ and $\beta^1=\beta^2=\beta^3$.
Then, the line element boils down to
\begin{equation}
ds^2 = -4N^2dt^2 + \frac1{N^2}(dz^2 + e^{2(\psi+\gamma)}dwd\bar w)\,,
\end{equation}
where
\begin{equation}
N^2 = \frac{(z^2+c)^2}{8(\alpha^0z+\beta^0)^{1/2}(\alpha^1z+\beta^1)^{3/2}}\,, \qquad e^{2\psi} =
(z^2+c)^2\,.
\end{equation}
Notice that $e^{2\psi}$ is still a quartic polynomial in $z$, but without a linear term. This suggests a
nonextremal generalization with $e^{2\psi}$ a generic quartic polynomial. In what follows, we shall
apply this idea to various prepotentials.

\subsection{The $F = -iX^0 X^1$ model}
\label{nonextr-X0X1}

Let us first consider the $\text{SU}(1,1)/\text{U}(1)$ model with prepotential
\[
	F(X) = -i X^0 X^1 \;.
\]
Supersymmetric solutions to this model were found in section 3.1 of \cite{Cacciatori:2009iz}.
 	
The Hesse potential for this model is given by (\ref{eq:H1}):
	\begin{align*}
			H(x,y) &= - 2 \Big( x^0 x^1 + y_0 y_1 \Big) \;. 
	\end{align*}
	We will consider axion-free solutions for which
	\begin{equation}
		x^0 = x^1 = 0 \;, \qquad \Rightarrow \qquad v_0 = v_1 = 0 \;. \label{eq:axion_free_condition3}
	\end{equation}
	We will further impose that ${\cal H}_0$ and ${\cal H}_1$ are constant, which means the electric charges vanish and we have a purely magnetic solution.
	The matrix $\tilde{H}^{ab}$ is given by
		\begin{equation}
		\tilde{H}^{ab} = \left( \begin{array}{cccc} 
		*	& * & 0 & 0 \\
		*	& * & 0 & 0 \\
		0 & 0 & 2{y_0}^2 & 0\\
		0 & 0 & 0 & 2{y_1}^2		
		\end{array} \right)\,. \label{eq:Htilde3}
	\end{equation}
The entries in the upper-left block can be easily computed, but are not relevant to our discussion since they completely decouple from the equations of motion for the class of solutions under consideration.
The dual coordinates, defined by $q_a := \partial_a \tilde{H} = -\tilde{H}_{ab} q^b$, are given by
	\[
		q_0 = 0 \;, \hspace{1.5em} q_1 = 0 \;, \hspace{1.5em} q_2 = -\frac{1}{2y_0} \;, \hspace{1.5em} q_3 = -\frac{1}{2y_1} \;.
	\]

Using the formula (\ref{eq:potential_axion_free_identity}) to compute $V(q)/H$, the equations of motion
\eqref{eq:eom1} become
	\begin{align}
		&\Delta {q}_2 - \frac{\Big[ (\partial_z q_2)^2 - (\partial_z {\cal H}_2)^2 \Big]}{q_2} + 4q_2 \left[(g_0 q_2)^2 + 2g_0 g_1 q_2 q_3\right] = 0\,, \label{eq:eom1a_axion_free3} \\
		&\Delta {q}_3 - \frac{\Big[ (\partial_z q_3)^2 - (\partial_z {\cal H}_3)^2 \Big]}{q_3} + 4q_3 \left[(g_1 q_3)^2 + 2g_0 g_1 q_2 q_3\right] = 0\,, \label{eq:eom1b_axion_free3}
	\end{align}	
and the Einstein equations \eqref{eq:eom3} boil down to
	\begin{align}
		& \frac{\Big[ (\partial_z q_2)^2 - (\partial_z {\cal H}_2)^2 \Big]}{2 q_2^2} + \frac{\Big[ (\partial_z q_3)^2 - (\partial_z {\cal H}_3)^2 \Big]}{2 q_3^2} \notag \\
		&\hspace{8em} - 2 \left[ \left(g_0 q_2 \right)^2 + \left(g_1 q_3 \right)^2 + 4 g_0 g_1 q_2 q_3 \right] = -\partial^2_z \psi - (\partial_z \psi)^2\,, \label{eq:eom2_axion_free3}
\\  		&4 \left[ \left(g_0 q_2 \right)^2 + \left(g_1 q_3 \right)^2 + 4 g_0 g_1 q_2 q_3 \right] = \partial^2_z \psi + 2(\partial_z \psi)^2 - \kappa e^{-2\psi}\,. \label{eq:eom3_axion_free3}
	\end{align}
We have not displayed the components of the Einstein equations where $\mu = z, \nu \neq z$ as they are
automatically satisfied once the separation ansatz \eqref{Phi-separable} holds.
Note that the fields ${\cal H}_2$ and ${\cal H}_3$ are harmonic functions, i.e.
	\[
		\partial_z {\cal H}_2 = A e^{-2\psi} \;, \hspace{2em} \partial_z {\cal H}_3 = B e^{-2\psi} \;,
	\]
	where $A$ and $B$ are some constants proportional to the magnetic charges.
	
We now wish to solve \eqref{eq:eom1a_axion_free3}-\eqref{eq:eom3_axion_free3}. To this end,
inspired by the BPS case \cite{Cacciatori:2009iz} and by the considerations at the beginning of this
section, we make the ansatz
\begin{equation}
q_2 = \frac{f_2}{e^{\psi}}\,, \qquad q_3 = \frac{f_3}{e^{\psi}}\,, \qquad f_2 = \alpha_2 z + \beta_2\,,
\qquad f_3 = \alpha_3 z + \beta_3\,, \label{ansatz-q2q3}
\end{equation}
where $\alpha_2,\alpha_3,\beta_2,\beta_3$ are constants, and $e^{2\psi}$ is a quartic polynomial,
\begin{equation}
e^{2\psi} = \sum_{n=0}^4 a_n z^n\,. \label{quartic}
\end{equation}
Notice that the four-dimensional geometry has two scaling symmetries, namely
\[
(t,z,w,e^{\phi},e^{\psi}) \mapsto (t/\mu,\mu z,\mu w,e^{\phi}\mu^2,e^{\psi})\,,
\]
and
\[
(t,z,w,e^{\phi},e^{\psi}) \mapsto (t/\mu,\mu z,w,e^{\phi}\mu^2,e^{\psi}\mu)\,.
\]
One can use the first to set $\kappa=0,\pm1$ (corresponding to $\mathbb{R}^2$, $\text{S}^2$ and
$\text{H}^2$ respectively) and then the second (that leaves $\kappa$ invariant) to choose $a_4=1$.
Furthermore, by shifting the coordinate $z$, it is always possible to eliminate the cubic term in \eqref{quartic}.
We shall thus take $a_3=0$ in what follows. After that, it is straightforward to verify that the equations
of motion \eqref{eq:eom1a_axion_free3}-\eqref{eq:eom3_axion_free3} are satisfied if the following
relations for the coefficients hold:
\[
\alpha_2 = \frac1{2g_0}\,, \qquad \alpha_3 = \frac1{2g_1}\,, \qquad \beta_2g_0 + \beta_3g_1 = 0\,,
\]
\[
a_0 = 2(A^2g_0^2 + B^2g_1^2) - 4\beta_2^2g_0^2(\kappa - 4\beta_2^2g_0^2)\,, \qquad
a_1 = \frac{B^2g_1^2 - A^2g_0^2}{\beta_2 g_0}\,,
\]
\[
a_2 = \kappa - 8\beta_2^2 g_0^2\,, \qquad a_3 = 0\,, \qquad a_4 = 1\,.
\]
We are thus left with a three-parameter family of solutions, labeled by $(A,B,\beta_2)$. Note that
the eqns.~\eqref{static}, \eqref{additional-ansatz} are trivially satisfied in this case. The dilaton
$\phi$ is computed from the Hesse potential (\ref{eq:Hesse_KK}),
\[
e^{\phi} = -2H(x,y) = \frac{e^{2\psi}}{f_2f_3}\,.
\]
Introducing coordinates $\vartheta,\varphi$ according to
\[
w = \left\{\begin{array}{r@{\quad,\quad}l} 2\tan\frac{\vartheta}2 e^{i\varphi} & \kappa = 1 \\
                                                                                    \vartheta e^{i\varphi} & \kappa = 0 \\
                                                                                    2\tanh\frac{\vartheta}2 e^{i\varphi} & \kappa = -1
                                                                                    \end{array}\right.
\]
yields for the four-dimensional metric
\begin{equation}
ds_4^2 = -\frac{e^{2\psi}}{f_2f_3}dt^2 + \frac{f_2f_3}{e^{2\psi}}dz^2 + f_2f_3(d\vartheta^2 +
S_{\kappa}^2(\vartheta)d\varphi^2)\,, \label{metric-X0X1}
\end{equation}
where we defined
\[
S_{\kappa}(\vartheta) = \left\{\begin{array}{c@{\quad,\quad}l} \sin\vartheta & \kappa = 1 \\
                                                                                    \vartheta & \kappa = 0 \\
                                                                                    \sinh\vartheta & \kappa = -1\,.
                                                                                    \end{array}\right.
\]
Moreover, one has from \eqref{xy}
\[
X^0 = -\frac{i}{2}\left(\frac{g_1}{g_0}\right)^{\frac12}\!\!\left(\frac{z + 2g_0\beta_2}{z - 2g_0\beta_2}
\right)^{\frac12}\,, \qquad X^1 = -\frac{i}{2}\left(\frac{g_0}{g_1}\right)^{\frac12}\!\!\left(\frac{z - 2g_0\beta_2}
{z + 2g_0\beta_2}\right)^{\frac12}\,.
\]
Finally, from (\ref{eq:field_strengths}) the gauge field strengths read
\begin{align*}
[F^0]_{\mu 0} &= 0 \;, &  [F^1]_{\mu 0} &= 0 \;, \\
[G_0]_{z 0} &= \frac{A}{f_2^2} \;, \hspace{1em}[G_0]_{i 0} = 0\;, & [G_1]_{z 0} &= \frac{B}{f_3^2} \;, \hspace{1em}[G_1]_{i 0} = 0\;,
\end{align*}
and using the fact that
\[
	{\cal N}_{IJ} = \left( \begin{array}{cc} -i\frac{f_3}{f_2} & 0 \\ 0 &  -i\frac{f_2}{f_3} \end{array}\right) \;,
\]
we can write this as
\[
F^0 = \frac i2 A e^{2\gamma} dw\wedge d\bar w\,, \hspace{2em} F^1 = 
\frac i2 B e^{2\gamma} dw\wedge d\bar w\,.
\]
Observe that the expressions for the gauge field strengths are precisely the same as for the BPS
case \cite{Cacciatori:2009iz}. The solution \eqref{metric-X0X1} has an event horizon at the largest
root $z_{\text h}$ of $e^{2\psi}=0$. Regularity of the Euclidean section at $z=z_{\text h}$ gives the
Hawking temperature
\[
T = \frac{(e^{2\psi})'|_{z_{\text h}}g_0g_1}{\pi(z^2_{\text h} - 4\beta_2^2g_0^2)}\,.
\]
For the Bekenstein-Hawking entropy one obtains
\begin{equation}
S = \frac{A_{\text h}}{4G} = \frac{V}{16Gg_0g_1}(z_{\text h}^2 - 4\beta_2^2g_0^2)\,,
\end{equation}
where\footnote{If the horizon is noncompact, one can still define a finite entropy density $s=S/V$.}
\eq
V \equiv \int S_{\kappa}(\vartheta)d\vartheta d\varphi\,. \label{def-vol}
\feq
The BPS solution found in \cite{Cacciatori:2009iz} is recovered for
\[
A g_0 = B g_1 = -\frac{\kappa}4\,.
\]
In the BPS case, the magnetic charges $A,B$ obey thus a Dirac-type quantization condition\footnote{The
magnetic charge densities $p^I$ of \cite{Cacciatori:2009iz} are related to $A,B$ by $A=4\pi p^0$,
$B=4\pi p^1$.}. Note that the supersymmetric solution describes a genuine black hole only for
$\kappa=-1$.

	
\subsection{The $t^3$ model}
\label{tcube}

This model is characterised by the prepotential
\[
	F(X) = -2i\sqrt{X^0 {X^1}^3} \;.
\]
 	
The Hesse potential corresponding to this model is given by (\ref{eq:H2}):
	\begin{align*}
			H(x,y) &= - 2 \Big( -y_0 x^0 y_0 x^0 + 2 y_0 x^0 y_1 x^1 + \frac{1}{3} y_1 x^1 y_1 x^1 + \frac{4}{27} y_0 (y_1)^3 + 4 x^0 (x^1)^3 \Big)^{1/2} \;. 
	\end{align*}
	We will consider axion-free solutions for which
	\begin{equation}
		x^0 = x^1 = 0 \;, \qquad \Rightarrow \qquad v_0 = v_1 = 0 \;. \label{eq:axion_free_condition4}
	\end{equation}
	We impose further that ${\cal H}_0$ and ${\cal H}_1$ are constant, so we are dealing with purely magnetic solutions.
	The matrix $\tilde{H}^{ab}$ reads
		\begin{equation}
		\tilde{H}^{ab} = \left( \begin{array}{cccc} 
		*	& * & 0 & 0 \\
		*	& * & 0 & 0 \\
		0 & 0 & 4{y_0}^2 & 0\\
		0 & 0 & 0 & \frac{4}{3}{y_1}^2		
		\end{array} \right) \;. \label{eq:Htilde4}
	\end{equation}
	The dual coordinates, defined by $q_a := \partial_a \tilde{H} = -\tilde{H}_{ab} q^b$, are given by
	\[
		q_0 = 0 \;, \hspace{1.5em} q_1 = 0 \;, \hspace{1.5em} q_2 = -\frac{1}{4y_0} \;, \hspace{1.5em} q_3 = -\frac{3}{4y_1} \;.
	\]
	
Using the formula (\ref{eq:potential_axion_free_identity}) for $V(q)/H$, the equations of motion
\eqref{eq:eom1} become
	\begin{align}
		&\Delta q_2 - \frac{\Big[ (\partial_z q_2)^2 - (\partial_z {\cal H}_2)^2 \Big]}{q_2} + 16 q_2 (g_0 g_1 q_2 q_3) = 0 \;, \label{eq:eom1a_axion_free4} \\
		&\Delta q_3 - \frac{\Big[ (\partial_z q_3)^2 - (\partial_z {\cal H}_3)^2 \Big]}{q_3} + q_3 \left[ \frac{16}{3}g_0 g_1 q_2 q_3 + \frac{32}{9}(g_1 q_3)^2  \right] = 0 \;, \label{eq:eom1b_axion_free4} 
	\end{align}
	and the Einstein equations \eqref{eq:eom3} are given by
	\begin{align}
		& \frac{\Big[ (\partial_z q_2)^2 - (\partial_z {\cal H}_2)^2 \Big]}{4 q_2^2} + \frac{3 \Big[ (\partial_z q_3)^2 - (\partial_z {\cal H}_3)^2 \Big]}{4 q_3^2} \notag \\
		&\hspace{8em} - 8g_0g_1q_2 q_3 - \frac{8}{3}(g_1 q_3)^2 = -\partial^2_z \psi - (\partial_z \psi)^2 \;, \label{eq:eom2_axion_free4} \\
	 		&16g_0g_1q_2 q_3 + \frac{16}{3}(g_1 q_3)^2 = \partial^2_z \psi + 2(\partial_z \psi)^2 - \kappa e^{-2\psi}\,,
			\label{eq:eom3_axion_free4}
	\end{align}	
Again, the Einstein equations with $\mu = z, \nu \neq z$ hold identically by virtue of \eqref{Phi-separable}. 
Note that ${\cal H}_2$ and ${\cal H}_3$ are harmonic functions, i.e.
	\[
		\partial_z {\cal H}_2 = A e^{-2\psi} \;, \hspace{2em} \partial_z {\cal H}_3 = B e^{-2\psi} \;,
	\]
where $A$ and $B$ are some constants proportional to the magnetic charges.

In order to find nonextremal black hole solutions to the $t^3$ model, we use again the ansatz
\eqref{ansatz-q2q3}, \eqref{quartic}. Without loss of generality,
we shall choose $\kappa=0,\pm 1$, $a_4=1$, $a_3=0$ by employing the various scaling and
shift symmetries explained in section \ref{nonextr-X0X1}. One finds then that the equations of
motion \eqref{eq:eom1a_axion_free4}-\eqref{eq:eom3_axion_free4} hold if the coefficients satisfy
\begin{align*}
\alpha_2 &= \frac1{4g_0}\,, \qquad \alpha_3 = \frac3{4g_1}\,, \qquad \beta_2g_0 + \beta_3g_1 = 0\,, \\
a_0 &= \frac43\left(3A^2g_0^2 + B^2g_1^2\right) - \frac{16}3\beta_2^2g_0^2\left(\frac{16}9
\beta_2^2g_0^2 + \kappa\right)\,, \\
a_1 &= \frac{B^2g_1^2 - 9A^2g_0^2}{3\beta_2g_0} + \frac83\beta_2g_0\left(\frac{64}9
\beta_2^2g_0^2 + \kappa\right)\,, \\
a_2 &= \kappa - \frac{32}3\beta_2^2 g_0^2\,, \qquad a_3 = 0\,, \qquad a_4 = 1\,,
\end{align*}
and thus we have again a three-parameter family of solutions, labeled by $(A,B,\beta_2)$.
As before, the eqns.~\eqref{static}, \eqref{additional-ansatz} are trivially satisfied in this case.
For the dilaton, one obtains
\[
e^{\phi} = -2H(x,y) =  \frac{e^{2\psi}}{2f_2^{1/2}f_3^{3/2}}\,,
\]
such that the four-dimensional metric reads
\begin{equation}
ds_4^2 = -\frac{e^{2\psi}}{2f_2^{1/2}f_3^{3/2}} dt^2 + \frac{2f_2^{1/2}f_3^{3/2}}{e^{2\psi}} dz^2 +
2f_2^{1/2}f_3^{3/2} \left(d\vartheta^2 + S_{\kappa}^2(\vartheta)d\varphi^2\right)\,. \label{metric-t^3}
\end{equation}
For the upper part of the symplectic section one has from \eqref{xy}
\[
X^0 = -i\left(\frac{g_1}{g_0}\right)^{\frac34}\!\!\left(\frac{z + 4g_0\beta_2}{12z - 16g_0\beta_2}
\right)^{\frac34}\,, \qquad X^1 = -\frac{i}{2}\left(\frac{g_0}{g_1}\right)^{\frac14}\!\!\left(\frac{3z - 4g_0\beta_2}
{4z + 16g_0\beta_2}\right)^{\frac14}\,.
\]
Finally, from (\ref{eq:field_strengths}) the gauge field strengths read
\begin{align*}
[F^0]_{\mu 0} &= 0 \;, &  [F^1]_{\mu 0} &= 0 \;, \\
[G_0]_{z 0} &= \frac{A}{2f_2^2} \;, \hspace{1em}[G_0]_{i 0} = 0\;, & [G_1]_{z 0} &= \frac{3B}{2f_3^2} \;, \hspace{1em}[G_1]_{i 0} = 0\;,
\end{align*}
and using the fact that
\[
	{\cal N}_{IJ} = \left( \begin{array}{cc} -i\left(\frac{f_3}{f_2}\right)^{3/2} & 0 \\ 0 &
	-3i\left(\frac{f_2}{f_3}\right)^{1/2} \end{array}\right)\,,
\]
we can write this as
\[
F^0 = \frac i2 A e^{2\gamma} dw\wedge d\bar w\,, \hspace{2em} F^1 = \frac i2 B e^{2\gamma} dw
\wedge d\bar w\,,
\]
which is the same as in the BPS case \cite{Cacciatori:2009iz}. The Hawking temperature and
entropy are given respectively by
\eq
T = \frac{2(e^{2\psi})'|_{z_{\text h}}g_0^{1/2}g_1^{3/2}}{\pi(z_{\text h} + 4g_0\beta_2)^{1/2}
(3z_{\text h} - 4g_0\beta_2)^{3/2}}\,,
\feq
\eq
S = \frac{V(z_{\text h} + 4g_0\beta_2)^{1/2}(3z_{\text h} - 4g_0\beta_2)^{3/2}}{32 G g_0^{1/2}g_1^{3/2}}\,,
\feq
where the horizon coordinate $z_{\text h}$ is the largest root of $e^{2\psi}=0$, and $V$ was defined
in \eqref{def-vol}.

The above nonextremal black hole boils down to the spherically symmetric BPS solution found
in \cite{Cacciatori:2009iz} in the special case
\[
\kappa = 1\,, \qquad Ag_0 = -\frac18 - \frac83\beta_2^2g_0^2\,, \qquad Bg_1 = -\frac38 + \frac83\beta_2^2g_0^2\,.
\]
Although the single charges $A,B$ are not quantized, the sum $Ag_0+Bg_1$ (which is equal to
$4\pi g_Ip^I$ in the notation of \cite{Cacciatori:2009iz}) is, because it is the linear combination
$g_IA^I$ that couples minimally to the gravitinos.

	
\subsection{The $F = -\frac{{X^1}^3}{X^0}$ model}

We now consider the model characterised by the prepotential
\eq
	F(X) = -\frac{{X^1}^3}{X^0}\,. \label{prepot-X1^3:X0}
\feq
 	
The Hesse potential corresponding to \eqref{prepot-X1^3:X0} is given by (\ref{eq:H3}):
	\begin{align*}
			H(x,y) &= - 2 \Big( -y_0 x^0 y_0 x^0 - 2 y_0 x^0 y_1 x^1 + \frac{1}{3} y_1 x^1 y_1 x^1 + \frac{4}{27} x^0 (y_1)^3 - 4 y_0 (x^1)^3 \Big)^{1/2} \;. 
	\end{align*}
	We will consider axion-free solutions which take the form
	\begin{equation}
		y_0 = x^1 = 0 \;, \qquad \Rightarrow \qquad u^0 = v_1 = 0 \;. \label{eq:axion_free_condition5}
	\end{equation}
	We further impose that ${\cal H}_1$ and ${\cal H}_2$ are constant, and we are left with one non-constant electric potential and one non-constant magnetic potential, corresponding to ${\cal H}_0$ and ${\cal H}_3$ respectively.
	The matrix $\tilde{H}^{ab}$ reads
		\begin{equation}
		\tilde{H}^{ab} = \left( \begin{array}{cccc} 
		4{x^0}^2	& 0 & 0 & 0 \\
		0	& * & * & 0 \\
		0 & * & * & 0\\
		0 & 0 & 0 & \frac{4}{3}{y_1}^2		
		\end{array} \right) \;. \label{eq:Htilde5}
	\end{equation}
	The dual coordinates, defined by $q_a := \partial_a \tilde{H} = -\tilde{H}_{ab} q^b$, are given by
	\[
		q_0 = -\frac{1}{4x^0} \;, \hspace{1.5em} q_1 = 0 \;, \hspace{1.5em} q_2 = 0 \;, \hspace{1.5em} q_3 = -\frac{3}{4y_1} \;.
	\]
	
Using the formula (\ref{eq:potential_H}) for $V(q)/H$, the equations of motion
\eqref{eq:eom1} become
 	\begin{align}
		&\Delta q_0 - \frac{\Big[ (\partial_z q_0)^2 - (\partial_z {\cal H}_0)^2 \Big]}{q_0} = 0 \;, \label{eq:eom1a_axion_free5} \\
		&\Delta q_3 - \frac{\Big[ (\partial_z q_3)^2 - (\partial_z {\cal H}_3)^2 \Big]}{q_3} + \frac{32}{9} g_1^2 q_3^3 = 0 \;, \label{eq:eom1b_axion_free5} 
	\end{align}
and the Einstein equations \eqref{eq:eom3} boil down to
	\begin{align}
		& \frac{\Big[ (\partial_z q_0)^2 - (\partial_z {\cal H}_0)^2 \Big]}{4 q_0^2} + \frac{3 \Big[ (\partial_z q_3)^2 - (\partial_z {\cal H}_3)^2 \Big]}{4 q_3^2} -\frac83 \left(g_1 q_3\right)^2 = -\partial^2_z \psi - (\partial_z \psi)^2\,,\label{eq:eom2_axion_free5} \\
	 		&\frac{16}3 \left(g_1 q_3 \right)^2 = \partial^2_z \psi + 2(\partial_z \psi)^2 - \kappa e^{-2\psi}\,.
			\label{eq:eom3_axion_free5}
	\end{align}	
As before,	 the Einstein equations with $\mu = z, \nu \neq z$ hold identically due to \eqref{Phi-separable}.
Note that ${\cal H}_0$ and ${\cal H}_3$ are harmonic functions, i.e.
	\[
		\partial_z {\cal H}_0 = -A e^{-2\psi} \;, \hspace{2em} \partial_z {\cal H}_3 = B e^{-2\psi} \;,
	\]
where $A$ and $B$ are some constants, with A proportional to the electric charge and $B$ proportional to the magnetic charge.

If we define a new coordinate $\zeta$ by $d\zeta=e^{-2\psi}dz$, \eqref{eq:eom1a_axion_free5}
reduces to the Liouville equation
\eq
\frac{d^2}{d\zeta^2}\ln q_0 = -\frac{A^2}{q_0^2}\,, \label{liou-q0}
\feq
whose solution reads
\[
q_0 = A\frac{\sin p\zeta}p\,,
\]
where $p$ (with $p^2$ real) is an integration constant, the so-called Liouville momentum\footnote{$p$
real or imaginary corresponds to the hyperbolic or elliptic solution respectively. The limiting case
$p=0$ is the parabolic solution.}. One has then
\[
\frac1{4q_0^2}\left[(\partial_z q_0)^2 - A^2e^{-4\psi}\right] = \frac{p^2}4e^{-4\psi}\,.
\]
Using this in \eqref{eq:eom2_axion_free5}, we see that $q_0$ decouples from the other
fields\footnote{The deeper reason why this happens is that the scalar potential is independent of
$g_0$ (even without imposing an axion-free condition).}. Unfortunately, the remaining eqns.~for
$q_3$ and $\psi$ are not solved by the ansatz \eqref{ansatz-q2q3}, \eqref{quartic}. The reason
for this is that, unlike the two models considered before, the scalar potential corresponding to
\eqref{prepot-X1^3:X0} has no critical point (it is just of Liouville-type), and thus there is no
AdS vacuum to which the black hole asymptotes. Already for the BPS solution \cite{Cacciatori:2009iz},
$e^{2\psi}$ is a complicated transcendental function and not a quartic polynomial. The latter case
has quantized magnetic charge, $4B^2g_1^2=\kappa^2$, as well as zero Liouville momentum, and thus
the first term in \eqref{eq:eom2_axion_free5} is absent. Then, the remaining
eqns.~\eqref{eq:eom1b_axion_free5}, \eqref{eq:eom2_axion_free5},
\eqref{eq:eom3_axion_free5} arise from the first-order system
\eq
\frac d{d\zeta}\varphi^{\alpha} = {\cal G}^{\alpha\beta}\frac{\partial W}{\partial\varphi^{\beta}}\,,
\label{1storder}
\feq
where $\alpha,\beta=1,2$, $\varphi^1\equiv\ln q_3$, $\varphi^2 \equiv\psi$, $({\cal G}^{\alpha\beta})
\equiv\text{diag}(4/3,-1)$, and the superpotential $W$ is given by
\eq
W = g_1e^{\varphi^1 + 2\varphi^2} + \frac{3\kappa}{8g_1}e^{-\varphi^1}\,.
\feq
\eqref{1storder} can easily be integrated to give
\[
q_3 = \left(Ce^{-4\psi/3} - \frac{3\kappa}{4g_1^2}e^{-2\psi}\right)^{1/2}\,,
\]
which is (3.77) of \cite{Cacciatori:2009iz} ($C$ denotes an integration constant). Using $\psi$ in place
of $z$ as a radial coordinate, one can then proceed to obtain the supersymmetric solution in section 3.3
of \cite{Cacciatori:2009iz}.

While it was to be expected that the BPS case follows from a set of first-order equations, it is rather
surprising that also some nonextremal black holes arise from a first-order system via a superpotential
construction, like e.g.~the Reissner-Nordstr\"om-AdS solution in any dimension \cite{Lu:2003iv}.
Given the results of \cite{Lu:2003iv} (cf.~also \cite{Miller:2006ay,Janssen:2007rc,Cardoso:2008gm,Andrianopoli:2009je,Galli:2011fq}
for related work), it would be very interesting to see if a class of
nonextremal black holes follows from first-order equations also for the $F=-(X^1)^3/X^0$ model
(and for the ones in sections \ref{nonextr-X0X1}, \ref{tcube}).
We shall come back to this point in a future publication \cite{to-appear}.

\section{Conclusions and final remarks}
\label{conclusions}

In this paper we constructed new finite temperature black hole solutions to Fayet-Iliopoulos gauged
matter-coupled supergravity. This was done for some simple prepotentials.
The generalization to more complicated models like the stu model with prepotential
$F=-2i(X^0X^1X^2X^3)^{1/2}$ is immediate and will be presented in a forthcoming paper \cite{to-appear}.
This solution will include both the BPS black holes found in  \cite{Cacciatori:2009iz} and the nonextremal
black holes of \cite{Duff:1999gh} with four magnetic charges\footnote{Note that the latter do not admit
a supersymmetric limit \cite{Duff:1999gh}.}.

It was found in \cite{Cvetic:2010mn} that for a large class of rotating multi-charge black holes in
asymptotically anti-de~Sitter spacetimes, the product of all horizon areas (including thus also inner
horizons) depends only on the charges, angular momenta and the cosmological constant.
It would be very interesting to see whether such universal results, which may provide a ``looking glass'' 
for probing the microscopics of general black holes, hold also for the solutions constructed here.

A further question is related to the geometry of the three-dimensional base space.
In ${\cal N}=2$, $D=4$ ungauged supergravity the most general static, spherically symmetric 
three-dimensional line element is the three-dimensional part of the Reissner-Nordstr\"om
metric \cite{Breitenlohner:1987dg} 
\eq
ds_3^2 = \frac{r_0^4}{\sinh^4 r_0\tau}d\tau^2 + \frac{r_0^2}{\sinh^2 r_0\tau}(d\vartheta^2 +
\sin^2\vartheta d\varphi^2)\,, \label{FGK-base}
\feq
where $\tau$ is a radial coordinate and $r_0$ denotes the nonextremality parameter.
(For $r_0\to 0$, $ds_3^2$ becomes the flat metric). It is therefore natural to suspect that in the nonextremal case the
flat three-dimensional base space of the extremal solutions is replaced by the three-dimensional 
part of the Reissner-Nordstr\"om solution. This was a key feature in a recent conjecture for deforming 
extremal into nonextremal black holes presented in \cite{Galli:2011fq}\footnote{For 
higher-dimensional generalizations see \cite{Meessen:2011bd}.}, which makes critical use of the formalism 
of Ferrara, Gibbons and Kallosh \cite{Ferrara:1997tw}. At present there is no proof of the conjecture of \cite{Galli:2011fq} for generic 
models, but it works for at least the stu model. One may therfore wonder if there is a similar 
underlying structure in gauged supergravity as well.
At first sight, the answer seems to be negative, since in the gauged case the base space depends on
the charges, whereas \eqref{FGK-base} is independent of them, but perhaps this issue is more
subtle than one might think at first sight. We hope to come back to these points in future work.


\acknowledgments

This work was partially supported by INFN and MIUR-PRIN contract 2009-KHZKRX. The work of O.V. is supported by an STFC studentship and by DAAD. O.V. would like to thank Thomas Mohaupt for useful discussions. We thank Thomas Van Riet for helpful comments on the first version of this paper.

\appendix

\section{Formula for the potential $V$ in terms of the Hesse potential}

The following is true for generic models. The matrix $\hat{H}_{ab}$ is defined to be
\[
	\hat{H}_{ab} := \left( \begin{array}{cc} {\cal I} + {\cal R} {\cal I}^{-1} {\cal R} & -{\cal R} {\cal I}^{-1} \\ -{\cal I}^{-1} {\cal R} & {\cal I}^{-1} \end{array} \right) \;.
\]
In equation (19) of \cite{Mohaupt:2011aa} this was shown to be related to $\tilde{H}_{ab}$ through the expression
\[
	\frac{1}{H} \hat{H}_{ab} = \tilde{H}_{ab} + \frac{2}{H^2} \Omega_{ac} q^c \Omega_{bd} q^d \;,
\]
where $\tilde{H} := -\frac12 \log(-2H)$ and $\tilde{H}_{ab} = \partial^2_{a,b} \tilde{H}$.
Restricting the indices in this expression to $a = I + n$ and $b = I + n$, where $I,J = 0, \ldots, n-1$ we have the expression
\begin{equation}
	\frac{1}{H} {\cal I}^{IJ} = \frac{\partial^2}{\partial y_I \partial y_J} \tilde{H} + \frac{2}{H^2} x^I x^J \;. \label{eq:IH_identity}
\end{equation}

The potential $V$ is given by
\[
V = -2g_I g_J \left({\cal I}^{IJ} - \frac{4}{H} \bar{X}^I X^J \right) =
       -2g_I g_J \left({\cal I}^{IJ} - \frac{4}{H} \left(x^I x^J + u^I u^J\right) \right)\,.
\]
Using the identity (\ref{eq:IH_identity}) we can write this as
\[
	\frac{1}{H}V = -2 g_I g_J \left(\frac{\partial^2}{\partial y_I \partial y_J} \tilde{H} - \frac{2}{H^2} x^I x^J - \frac{4}{H^2} u^I u^J \right) \;.
\]
Given the fact that
\[
	\frac{u^I}{H} = \frac{\partial \tilde{H}}{\partial y_I} \;,
\]
we can write this in terms of the Hesse potential as
\begin{equation}
	\frac{1}{H}V = 2 g_I g_J \left( -\frac{\partial^2}{\partial y_I \partial y_J} \tilde{H} + \frac{2}{H^2} x^I x^J + 4 \frac{\partial \tilde{H}}{\partial y_I} \frac{\partial \tilde{H}}{\partial y_J}\right) \;.
	\label{eq:potential_H}
\end{equation}

\subsection{Potential for axion-free solutions}

Often we will restrict to axion-free solutions, in which case
\[
	X^I = iu^I \;,
\]
i.e. $x^I = 0$.
The potential can then be written simply as
\begin{equation}
	\frac{1}{H}V = 2 g_I g_J \left(-\frac{\partial^2}{\partial y_I \partial y_J} \tilde{H} + 4 \frac{\partial \tilde{H}}{\partial y_I} \frac{\partial \tilde{H}}{\partial y_J}\right) \;.
	\label{eq:potential_axion_free_identity}
\end{equation}

\section{Hesse potentials}

Throughout this section, as in the main text, we adopt the notation
\begin{align*}
	x^I + iu^I = Y^I \;, \qquad y_I + i v_I = F_I \;.
\end{align*}

\subsection{Hesse potential for $F = -iY^0 Y^1$ model}

Taking derivatives of $F$ we find
\[
	F_0 = -iY^1\;, \qquad F_1 = -i Y^0\;.
\]
which implies immediately that
\begin{align*}
	v_0 &= -x^1 \;, \qquad u^1 = y_0 \;, \\
	v_1 &= -x^0 \;, \qquad u^0 = y_1 \;.
\end{align*}
The Hesse potential is a homogeneous function of degree two with respect to $q^a = (x^I, y_I)^T$, and, hence, can be written as
\[
	2H = q^a \partial_a H \;.
\]
Using the fact that for all models $\partial_a H = (2v_I, -2u^I)^T$ we find that the Hesse potential for this model is given simply by
\begin{equation}
		H(x,y) = -2\Big( x^0 x^1 + y_0 y_1 \Big) \;.
	\label{eq:H1}
\end{equation}

\subsection{Hesse potential for $t^3$ model}

	The $t^3$ model is characterised by the prepotential
	\[
		F(Y) = -2i\sqrt{Y^0 {Y^1}^3} \;.
	\]
	The K\"ahler potential for this model can be written as 
	\[
		e^{\cal -K} = -\bar{Y} N Y = 8 Y^0 \bar{Y}^0 \left[ \text{Im}(i\sqrt{Z^1}) \right]^3 \;,
	\]
	where $Z^1 := Y^1/Y^0$.	Since the Hesse potential is related to the K\"ahler potential through $e^{\cal -K} = -2H$, we are left to find expressions for $Y^0,Z^1$ in terms of $x^I,y_I$.

	Firstly, by direct calculation one can show that 
	\[
		\frac{y_1Z^1 - 3 y_0}{x^0 Z^1 - x^1} = 3i\sqrt{\bar{Z}^1} \;, \qquad \frac{3y_1 i\sqrt{Z^1} - 9x^1}{3x^0i \sqrt{Z^1} + y_1}  = 3i\sqrt{\bar{Z}^1} \;,
	\]
	which can be combined into the quadratic equation 
	\[
		\underbrace{\left(\tfrac{1}{9}y_1^2 + x^0 x^1\right)}_{a}\left(i\sqrt{Z^1}\right)^2 + \underbrace{\left(y_0 x^0 - \tfrac{1}{3} y_1 x^1\right)}_{b}i\sqrt{Z^1} + \underbrace{\left(\tfrac{1}{3} y_0 y_1 + {x^1}^2\right)}_{c} = 0 \;,
	\]
	with solution
	\[
		i\sqrt{Z^1} = \frac{-b \pm i \sqrt{4ac - b^2}}{2a} \;.
	\]
	In order to ensure that the K\"ahler potential is positive definite (and the Hesse potential negative definite) we make the positive sign choice in the above expression for $i\sqrt{Z^1}$.
	
	Again by direct calculation, one finds the following expression for $\bar{X}^0$:
	\[
		\bar{X}^0 = -i \frac{(x^0i\sqrt{Z^1} + \tfrac{1}{3}y_1)}{ \text{Im} \left( i\sqrt{Z^1} \right) } \;.
	\]
	Using the fact that
	\[
		\Big| x^0i\sqrt{Z^1}  + \tfrac{1}{3}y_1 \Big|^2 = a \;,
	\]
	the K\"ahler potential can be written as
	\[
		e^{\cal -K} = 4 \sqrt{4ac - b^2}  \;,
	\]
	and, hence,
	\begin{equation}
			H(x,y) = - 2 \Big( -y_0 x^0 y_0 x^0 + 2y_0 x^0 y_1 x^1 + \frac{1}{3} y_1 x^1 y_1 x^1 + \frac{4}{27} y_0 (y_1)^3 + 4 x^0 (x^1)^3 \Big)^{\frac{1}{2}} \;.
		\label{eq:H2}
	\end{equation}

\subsection{Hesse potential for $F = -\frac{{Y^1}^3}{Y^0}$ model}

	The K\"ahler potential for this model can be written as 
	\[
		e^{\cal -K} = -\bar{Y} N Y = 8 Y^0 \bar{Y}^0 \left[ \text{Im}(Z^1) \right]^3 \;,
	\]
	where again $Z^1 := Y^1/Y^0$. As in the case of the $t^3$ model, we use the fact that the Hesse potential is related to the K\"ahler potential through $e^{\cal -K} = -2H$, and we are left to find expressions for $Y^0,Z^1$ in terms of $x^I,y_I$.

	Firstly, by direct calculation one can show that 
	\[
		\dfrac{\frac13y_1Z^1 + y_0}{x^0 Z^1 - x^1} = -\bar{Z}^2 \;, \qquad \dfrac{x^1 Z^1 + \frac13 y_1}{x^0 Z^1 - x^1}  = \bar{Z}^1 \;,
	\]
	which can be combined into the quadratic equation 
	\[
		\underbrace{\left({x^1}^2 + \tfrac{1}{3} y_1 x^0\right)}_{a} \left(Z^1\right)^2 + \underbrace{\left(y_0 x^0 + \tfrac{1}{3} y_1 x^1\right)}_{b}Z^1 + \underbrace{\left(\tfrac{1}{9}{y_1}^2 - y_0 x^1\right)}_{c} = 0 \;,
	\]
	with solution
	\[
		Z^1 = \frac{-b \pm i \sqrt{4ac - b^2}}{2a} \;.
	\]
	In order to ensure that the K\"ahler potential is positive definite we make the positive sign choice in the above expression for $Z^1$.
	
	Again by direct calculation, one finds the following expression for $\bar{X}^0$:
	\[
		\bar{X}^0 = -i\frac{(x^0 Z^1 - x^1)}{ \text{Im} \left( Z^1 \right) } \;.
	\]
	Using the fact that
	\[
		\Big| x^0 Z^1 - x^1\Big|^2 = a \;,
	\]
	the K\"ahler potential can be written as
	\[
		e^{\cal -K} = 4 \sqrt{4ac - b^2}  \;,
	\]
	and, hence,
	\begin{equation}
			H(x,y) = - 2 \Big( -y_0 x^0 y_0 x^0 - 2y_0 x^0 y_1 x^1 + \frac{1}{3} y_1 x^1 y_1 x^1 + \frac{4}{27} x^0 (y_1)^3 - 4 y_0 (x^1)^3 \Big)^{\frac{1}{2}} \;.
		\label{eq:H3}
	\end{equation}


\end{document}